\documentclass[prb,twocolumn,showpacs,floatfix,amsfonts,superscriptaddress]{revtex4}
\usepackage{graphicx,graphics,color,epsfig}
\usepackage{bm}
\usepackage{amsmath}
\usepackage{amssymb}

\newcommand{\beqa}{\begin{eqnarray}}
\newcommand{\eeqa}{\end{eqnarray}}

\begin{document}
\preprint{}
\title{Non-uniform glassy
electronic phases from  competing local orders}
\author{Z. Nussinov}
    \affiliation{Dept of Physics, Washington University, St. Louis, MO
63160}
\author{I. Vekhter}
    \affiliation{Dept. of Physics and
Astronomy, Louisiana State University, Baton Rouge, LA 70803-4001}
\author{A. V. Balatsky}
    \affiliation{Theoretical Division, Los Alamos National
Laboratory, Los Alamos, New Mexico 87545}
\begin{abstract}
We study non-uniform states
and possible glassiness triggered by a competition
between distinct local orders in disorder free
systems. Both in Ginzburg-Landau theories and
in simple field theories, such inhomogeneous states arise from
negative gradient terms between the competing order
parameters. We discuss applications of these ideas to a variety
of strongly correlated systems.

\end{abstract}
 \maketitle

\section{Introduction.}
Accumulated experimental evidence strongly suggests that in many
correlated electronic systems, different types of ordering
phenomena compete and coexist over a wide range of tunable
parameters. The most ubiquitous such cohabitation is between
magnetic and superconducting orders. Itinerant antiferromagnetism
(AFM) coexists with superconductivity in the 115 heavy fermion
series (CeMIn$_5$, where $M=$Co,Ir, or In)\cite{Zapf}. In
$UPt_{3}$, superconductivity emerges at $T_{c} \approx 0.5 K$ from
a strongly correlated heavy electron state with small moment AFM
below 6K \cite{upt3}. In some of the high-T$_c$ cuprates
charge density order coexists with spin density order (``stripes''
\cite{jan,steve,tranquada}) and may be relevant to the onset of
the superconductivity and quantum critical behavior
\cite{review,dirk}. Recent measurements indicate that in
URu$_2$Si$_2$ there is a proliferation of competing phases under
an applied magnetic field \cite{KHKim:2003}. Experiments also
suggest multiple phases in the skutterudite superconductor
PrOs$_4$Sb$_{12}$ \cite{Yuji:PrOsSb}, manganites \cite{salamon},
and a number of other materials. Two trends are common to
these experimental findings. First, the
coexistence of different orders is often inhomogeneous. Second,
this coexistence is frequently most pronounced near a Quantum
Critical Point (QCP), where
the transition temperature for one of the order parameters
vanishes \cite{review, subir}.

Additionally, dynamics of compounds with inhomogeneous
coexistence of distinct orders is often glassy
\cite{BSimovic:2003,BSimovic:2004,TPark:2005,CPanagopoulos:2005}.
In some systems, such as manganites \cite{dagotto05} the glassy
behavior is, most likely, due to disorder upon doping the system.
In others, including cuprates, the glassiness may be
self-generated (not simply due to doping disorder
\cite{VMitrovic:2008}), and arise out of competing interactions at
different length scales \cite{jorg}. The question remains,
however, whether inhomogeneous and/or glassy behavior can arise
out of a theory with local interactions and no disorder. In this
article we address this question for a class of Ginzburg-Landau
theories with competing order parameters. A comprehensive survey
of classical systems with frustration and no disorder that display
glassy behavior and proliferation of inhomogeneous ground states,
can be found in Refs.~\onlinecite{les1, les2}.

We study a minimal Ginzburg-Landau (GL) \cite{tol} theory which
includes amplitude-gradient coupling between two distinct local
orders, and find the conditions for resultant inhomogeneous
phases. A related interesting work examining gradient
coupling in GL theories \cite{Mi} appeared slightly after the
initial dissemination of our results \cite{oldus}. Extensions of
the GL gradient couplings considered here are found in some
studies of the supersolid transition \cite{arun}. We show that,
for a range of parameters described below, our theory maps
onto an effective model that is likely to exhibit
glassiness. Whether a particular system does or does not show
glassy behavior upon cooling depends on the rate of temperature
change and other dynamical variables that are not part of our
equilibrium analysis. However, our approach allows us to conclude
whether a glassy phase is possible and likely to occur. In this we
follow the established approaches in the field \cite{jorg}.

The mapping that strongly suggests glassiness in our
approach is to a Brazovskii-like model for one of the order
parameters. The Brazovskii model \cite{Brazovskii} for a single
component order parameter is defined by a GL functional of the
form
\begin{eqnarray}
{\cal{F}} = \frac{V}{(2 \pi)^{d}} \int d^{d} k [\frac{r_{0}}{2}
+ D(|\vec{k}|- q)^{2})] |\Phi_{k}|^{2} + ...,
\label{br}
\end{eqnarray}
in momentum ($k$) space with $V$ the volume of the system. In
Eq.(\ref{br}), the ellipses denote cubic, quartic, and higher
order terms in the order parameter field $\Phi$. As the mass
term, $r_0$ changes sign, the transition to a broken symmetry
state $\Phi\neq 0$,involves the appearance of structures
characterized by a finite wavenumber on a shell of radius $q>0$).
Structures that satisfy definite commensurability relations
amongst the wavenumbers are most preferred. In
Ref.~\onlinecite{Brazovskii} Brazovskii found that large phase
space available for fluctuations around the minimizing shell
alters the character of the transition to the ordered state once
the fluctuations are accounted for, and suggested that it
becomes first order. Thermal fluctuations renormalize the cubic
terms of the GL theory. More recent replica calculations
\cite{jorg,glass,loh,DMFT} showed that the model has extensive
configurational entropy, indicating proliferation of modulated
low-energy states, and strongly suggesting slow dynamics and
glassiness under generic experimental conditions. Once again,
these replica calculations only establish that glassiness is a
plausible and likely alternative to the first order transition
into a uniformly modulated phase. Whether a finite temperature
Brazovskii transition does or does not transpire before the system
undergoes a dynamical arrest (the glass transition outlined below)
depends on microscopic details of the model. The known theoretical
techniques (SCSA, DMFT, and others) do not enable the proof of a
glassy phase. These methods only enable us to determine whether a
glassy phase is possible. \cite{glass, jorg, DMFT}

Below we find the mapping of systems with competing orders
to Brazovskii type models.  This mapping allows us to (i) Find
resultant inhomogeneous phases in the GL analysis; (ii) Include
fluctuations via a self-consistent field theory to establish that
one of the two scenarios is realized: (a) the critical temperature
for the onset of non-uniform states is suppressed to zero,
suggesting that these states are more likely to be observed near a
QCP; or, alternatively, (b) fluctuations lead to a low temperature
Brazovskii transition; (iii) Appeal to existing replica
calculation results to confirm the extensive configurational
entropy associated with these incommensurate structures in
disorder free systems with competing local orders, which strongly
suggests slow dynamics and glassiness. Finally, we comment on
possible realizations of our model and applicability of the
results to itinerant electronic systems.

\section{Ginzburg-Landau theory: instability
of uniform coexistence.} To empirically account for competing
orders, we analyze the Ginzburg-Landau (GL) functional with two
order parameters, $\Phi_1$ and $\Phi_2$, which we will choose to
be real and scalar without loss of generality.  We remark
that our very general GL approach applies to various types of order
parameters. Of course, the symmetry, number of components of the
order parameters etc. changes. Nevertheless, the conclusions are
generally much the same. The uniform part of the free energy is
${\cal F}_0=\int d{\bf x} F_0$ where
\begin{eqnarray}
  F_0=\frac{r_1}{2} |\Phi_1|^2 +  \frac{r_2}{2} |\Phi_2|^2 +
  \frac{t}{2} |\Phi_1|^2 |\Phi_2|^2 +\frac{1}{4}|\Phi_1|^4 +
    \frac{u}{4}|\Phi_2|^4.
\label{f0}
\end{eqnarray}
In the spirit of the GL theory, $r_{1,2}= a_{1,2} (T-T_{1,2})$,
with $T_{i}$ the mean field transition temperatures. All other
coefficients are taken to be temperature-independent. The
quadratic coupling of the order parameters is allowed for all
symmetries. We consider competing orders, $t>0$, so that the
uniform coexistence region ($\Phi_1\neq 0, \Phi_2\neq 0$) occurs
only below the lower of the transition temperatures, and for
$u>t_1^2$. In that case, the values of the fields minimizing the
free energy are $\widetilde\Phi_1^2=(r_2t-r_1 u)/(u-t^2)$, and
$\widetilde\Phi_2^2=(r_1t-r_2)/(u-t^2)$. In disorder free systems,
the only alternative to the uniform coexistence is phase
separation unless non-trivial gradient terms are present
\cite{Pryadko}. Therefore we include the inhomogeneous
contribution to the free energy, ${\cal F}_q=\int d{\bf x}
F_q({\bf x})$, where
\begin{eqnarray}
  F_q = \sum_{i} |\nabla\Phi_i|^2
- \sum_{i,j} g_{ij}|\Phi_i|^2 |\nabla \Phi_j|^2  + \sum_{i} p_i
|\nabla^2 \Phi_i|^2 . \label{fq}
\end{eqnarray}
Here, we included general symmetry allowed low order gradient
terms. To flesh out the quintessential physics in what follows, we
set $g_{11}=g_{22}=g_{21}=0,g_{12}>0$, and $p_1=0$. This is the essential
aspect of the model that allows us to investigate the appearance
of the inhomogeneous states. The coupling of the form
$-g_{12}|\Phi_1|^2 |\nabla \Phi_2|^2$ implies that in the effective
theory for the order parameter $\Phi_2$ the coefficient of the
gradient term, $1-g_{12}|\Phi_1|^2$, may become negative, making the
transition of the Brazovskii type. We now investigate when this is
possible.

\begin{figure}
\includegraphics[width=7cm]{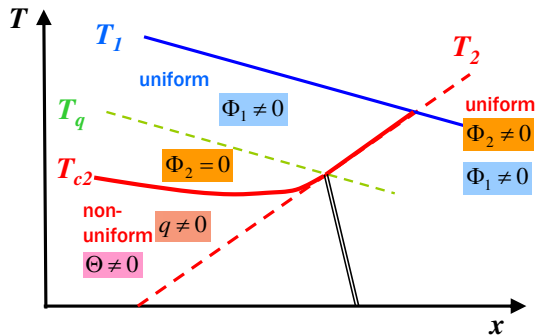}
\caption{The phase diagram obtained from the Ginzburg-Landau
expansion. The lines $T_{1,2}$ denote the bare mean field
transition temperatures as a function of a tuning parameter $x$.
$T_q$ is defined in text. An inhomogeneous phase appears below
$T_{c2}$. Double line denotes the first order transition.}
\label{Fig:GLPlot}
\end{figure}

With $F({\bf x}) = F_{0} + F_{q}$, the order parameter profiles
satisfy the Euler-Lagrange equations, $[\nabla \cdot (\partial
F/\partial (\nabla \Phi_{i}))] = (\partial F/\partial \Phi_{i})$.
By constructing inhomogeneous variational states whose free energy
is lower than the minimum amongst all possible uniform
configurations, we prove that the uniform solution is unstable
towards the appearance of inhomogeneities. We study the phase
diagram of the model assuming that the mean field transition
temperatures $T_i$ can be tuned by an external parameter, $x$
(pressure, doping, magnetic field etc.), as shown in
Fig.\ref{Fig:GLPlot}, with $T_{1}(x)$ monotonically decreasing,
and $T_{2}(x)$ monotonically increasing. That is,
\begin{eqnarray}
T_{1} = T_{1}^{(0)} - a_{1} x, \nonumber
\\ T_{2} = T_{2}^{(0)} + a_{2} x,
\end{eqnarray}
with $T_{1,2}^{(0)}$ and $a_{1,2}$ positive constants.

We first concentrate on the region $T_1>T_2$. Upon lowering the
temperature, the first transition is into the uniform state with
$\Phi_2=0$ and $\Phi_{1}({\bf x}) =-r_1$. Consequently, below the
$T_{q} = T_{1} - 1/(g_{12}a_{1})$ the coefficient of the $|\nabla
\Phi_2|^2$ term becomes negative indicating the tendency towards
the development of an inhomogeneous $\Phi_{2}$ phase. The
structure of this modulation depends on the difference $T_q-T_2$.
If this difference is sufficiently large, it is disadvantageous to
create non-vanishing bulk average of $\Phi_2$. Local
``bubbles'' of the order may appear upon lowering $T$, but their
study is not our focus in the present work.

In order to make the connection with the slow dynamics and
Brazovskii transition, we study the onset of the periodically
modulated phase of the form $\Phi_2({\bf x})=\Theta_2\cos({\bm
q}_i\cdot {\bf  x})$. Of the numerous contending low (free) energy
configurations, we will focus on analytically tractable modulated
structures; we do so in order to obtain stringent variational
bounds that we are able to extremize, and based on the
original analysis that showed the single modulation structures are
most advantageous \cite{Brazovskii}. In the regime $T_q\geq T_2$
minimization of the GL functional with respect to both $q$ and
$\Theta_2$ gives the transition temperature
\begin{eqnarray}
    T_{c2}&=&T_q-(g_{12} a_1)^{-1}
    \biggl[\sqrt{z^2+\frac{2tp}{g_{12}}+2pa_2(T_q-T_2)}-z\biggr]
  , \nonumber
\\ z &\equiv& \frac{pa_2-tpa_1}{g_{12}a_1}
\end{eqnarray}
to the phase $\Theta_2\neq 0$  with modulations at a {\it finite}
wave vector,
\begin{eqnarray}
q = \sqrt{ \frac{g_{12} a_{1} (T_{1} - T_{c2})}{2p_{2}}}.
\end{eqnarray}

In the regime $T_q < T_{1,2}$ the first transition, at $T_{q} <
T$, occurs into a spatially homogeneous phase. We next investigate
the phase diagram for the more general variational ansatz
$\Phi_2^{var}({\bf x})=\overline{\Phi}_2+\Theta_2\cos({\bm
q}_i\cdot {\bf x})$. Introduction of spatial modulations reduces
the condensation energy and therefore is unfavorable, unless
compensated by a significant gain due to the negative gradient
term. As a result we find a (generically first order, but
dependent on the magnitude of the coefficients in the GL
expansion) transition from the homogeneous to modulated, with a
finite $q$, phase at low $T$. In Fig.(\ref{Fig:GLPlot}), we show
the phase diagram of
 Eqs.(\ref{f0})-(\ref{fq}) for $t=0$. Of course, since we allowed only for the
restricted variational states  in the above analysis, our bounds
are more potent for the global free energy minima - $\Phi_{2}$ is
strictly inhomogeneous for all $T<T_{c2}(x)$; unrestricted
inhomogeneous states (not bound to the form of $\Phi_2^{var}$) may
extend to temperatures somewhat higher than $T_{c2}(x)$.

\section{Self-consistent field theory for competing order parameters.}

To improve on the GL analysis and incorporate the effect of
fluctuations self-consistently, we generalize our model to
$n$-component vector fields and utilize a large $n$ expansion.
As well known, the $n=\infty$ limit is equivalent to the spherical model describing
single component (scalar) particles \cite{spherical}. The physical
engine for the inhomogeneities is, as in preceding section,  the
amplitude gradient coupling which drives non-uniformities in
$\Phi_{2}$ once $\Phi_{1}$ is finite. For a finite
$\Phi_{1}({\bf{x}})= \Phi_{1}$, the effective free energy for
$\Phi_{2}$ is
\begin{eqnarray}
{\cal F}_{eff;2} &=& \int \frac{d^{d}k}{(2 \pi)^{d}} \Big[
(\frac{r_{2}}{2} + \frac{t}{2} \Phi_{1}^{2})
+ (1- g_{12} \Phi_{1}^{2}) k^{2} \nonumber
+ p k^{4} \big] \nonumber
\\ &\times&  \Phi_{2}({\bf k}) \Phi_{2}(-{\bf k})
+ \frac{u}{4}
\int \frac{d^{d}k_{1}}{(2 \pi)^{d}}
\frac{d^{d}k_{2}}{(2 \pi)^{d}}
\frac{d^{d}k_{3}}{(2 \pi)^{d}} \nonumber
\\ &\times& \Phi_{2}({\bf k}_{1}) \Phi_{2}({\bf k}_{2})
\Phi_{2}({\bf k}_{3}) \Phi_{2}(- {\bf k}_{1} - {\bf k}_{2} - {\bf
k}_{3})\,, \label{feff}
\end{eqnarray}
where $d$ is the dimensionality of the system.  The bare inverse
Green's functions are given by $G_{0}^{-1} = [r_{2}/2 + t/2 +
(1-g_{12} \Phi_{1}^{2}) k^{2} + pk^{4}]$. Incorporating
fluctuations self-consistently, we have $G^{-1} =
[{\overline{r}}_{2}/2 + (1-g_{12} \Phi_{1}^{2}) k^{2} + pk^{4}]$
where, by the Dyson equation, ${\overline{r}}_{2}/2 = r_{2}/2 +
t/2 + \Sigma$. To lowest order in $1/n$, the self-energy is given
by $\Sigma^{0} = \int \frac{d^{d}k}{(2 \pi)^{d}} G({\bf k})$, see
Ref.~ \onlinecite{Ma}. This leads to a self-consistency equation
for ${\overline{r}}_{2}$. Similar self-consistency equations
appear for $\Phi_{1}$; before the transition to an ordered
$\Phi_{2}$ state, $\Phi_{1}^{2} = - r_{1}$. A phase transition to
an ordered state $\Phi_{2} \neq 0$ occurs when the Green's
function acquires a pole on the real $k$ axis. If the pole is at
$k_{\min} =0$, the transition is to a uniform phase of $\Phi_{2}$;
if the pole first appears for $k_{\min} \neq 0$, the transition is
into a modulated phase.

When $[1- g_{12} \Phi_{1}^{2} ]>0$ the minimum of $G^{-1}$
is always at $k=0$, and both $\Phi_{1}$ and $\Phi_2$ may exhibit
uniform orders. On the other hand, if $[1- g_{12}
\phi_{1}^{2} ]<0$, the minimum for the $\Phi_{2}$ inverse Green's
function, $G^{-1}({\bf k})$ occurs at $k_{min} =  - [1- g_{12}
\Phi_{1}^{2} ]/(2p)$ leading to a real axis pole when
${\overline{r}}_{2}= {\overline{r}}_{2~\min} = [1- g_{12}
\Phi_{1}^{2} ]^{2}/(2p_{2})$. The quartic $G^{-1}$ has two
pairs of complex conjugate poles in the $k$ plane which lie on a
circle of radius $\rho = ({\overline{r}}_{2}/ (2
p_{12}))^{1/4}$. The finite real component of the poles
means that the correlation function $\langle \Phi_{2}({\bf x})
\Phi_{2}({\bf y}) \rangle$ exhibits sinusoidal modulations in
addition to exponential decay. The modulation and correlation
lengths are given, respectively, by
\begin{eqnarray}
l_{2} = 4 \pi [\sqrt{{\overline{r}}_{2}/2} +
 (1- g_{12} \Phi_{1}^{2}) /2]^{-1/2}, \nonumber
\\ \xi_{2} = 2[\sqrt{ {\overline{r}}_{2}/2} -(1- g_{12} \Phi_{1}^{2})/2]^{-1/2},
\label{G2}
\end{eqnarray}
with $\Phi_{1}$ the uniform competing order field.

Irrespective of the spatial dimensionality, whenever $[1- g_{12}
\Phi_{1}^{2} ]<0$, as ${\overline{r}}_{2} \to {\overline{r}}_{2~
\min}$ the self-energy diverges as $\Sigma \sim
({\overline{r}}_{2} - {\overline{r}}_{2~\min})^{-1/2}$. The phase
transition which would occur (at the mean field level) when
${\overline{r}}_{2} ={\overline{r}}_{2~\min}$ is thwarted by the
divergence of the self energy due to fluctuations. This
implies that $T_{c}=0$ similar to systems with competing
long range interactions \cite{us}. However, finite $n$
corrections (especially for $n=1$), may make the transition
temperature finite. In this case, this low temperature transition
for $\Phi_{2}$ is of Brazovskii type, with a shell of minimizing
modes.

A similar analysis holds for competing local orders in large $n$
quantum systems by extending \cite{us,glass}.  For bosonic fields,
after a summation over Matsubara frequencies, the $\Phi_{2}$
correlator is
\begin{eqnarray}
G_{2}({\bf k}) = \frac{\frac{1}{2} +
n_{B}\left(\sqrt{[\frac{{\overline{r}}_{2}} {2} + (1-g_{12}
\Phi_{1}^{2}) k^{2} + pk^{4}]/ (k_{B}T)}\right)}
{\sqrt{\frac{{\overline{r}}_{2}}{2} + (1-g_{12} \Phi_{1}^{2})
k^{2} + pk^{4}}}, \nonumber
\end{eqnarray} with
$n_{B}(x) = [\exp(x)-1]^{-1}$. Order, at large $n$, is still
inhibited in the quantum rendition of our system, although the
divergence of the self-energy in this case is less severe than for
its classical counterpart. In the bosonic system, due to
integration over imaginary time, the $\Phi_{2}$ self energy
diverges as $- \ln|{\overline{r}}_{2}- {\overline{r}}_{2~\min}|$
when ${\overline{r}}_{2} \to {\overline{r}}_{2~\min})$, whereas in
the classical system it diverges as $[{\overline{r}}_{2}-
{\overline{r}}_{2~\min}]^{-1/2}$.
Thermal fluctuations in classical $n=2$ systems
(e.g. complex scalar fields) also lead to a divergence of the $-
\ln|{\overline{r}}_{2}- {\overline{r}}_{2~\min}|$] type. \cite{glass, us, Schmalian}
The order probed by $\Phi_{2}$
is stabilized if the degeneracy of the minimizing wave-numbers is
lifted by augmenting the rotationally symmetric Hamiltonian by
additional lattice point group symmetry terms. In such instances,
the critical temperature of the large $n$ system often remains
anomalously low, and attains its minimal value exactly at the
onset of incommensurate order (e.g., when the minimizing modes are
of vanishing norm($q \to 0^{+}$)) \cite{us}. Degeneracy can also
be lifted by an external field which lowers the full rotational
symmetry of $H$ to a lower rotational symmetry in a plane
orthogonal to the field direction. For small $n$, a ``quantum''
finite temperature Brazovskii transition may occur for $\Phi_{2}$.

\section{Slow dynamics and glassiness.}

The key conclusion of the previous section was that our
model of competing orders, Eqs.(\ref{f0})-(\ref{fq}), maps onto a
Brazovskii-like model for the subdominant order parameter. The
transition temperature for the onset of this order is then
suppressed relative to the mean field conclusions due to a large
phase space available to low energy thermal and quantum
fluctuations. There are two possibilities for the transition
itself. It may take place as a first order fluctuation-induced
transition as originally envisioned by Brazovskii. Recent work on
a model equivalent to our Eq.~(\ref{feff}) suggests that a glass
transition may be realized as an alternative~\cite{jorg,loh}.

When applied to Eq.(\ref{feff}), the self-consistent screening
approximation shows \cite{jorg,loh} that the
configurational entropy, $S_c=k_B\log N_m$, with $N_m$  the number
of the metastable states, is extensive (proportional to the
volume) over a finite temperature range ($T_{A} > T
>T_{K}$), which depends on the coefficients of our GL expansion.
This entropy is due exclusively to the inhomogeneous field
$\Phi_{2}({\bf x})$ triggered by the competing uniform order
$\Phi_{1}({\bf x}) = \Phi_{1}$. At the onset ($T=T_{A}$)
\begin{eqnarray}
S_c(T_{A}) \approx C k_{B} (g_{12} \Phi_{1}-1)^{3} V,
\end{eqnarray}
where, in 3D, the numerical constant $C \simeq 1.18 \times
10^{-3}$ and $V$ is the volume \cite{jorg}. Our resulting
effective model for the inhomogeneous order is exactly equivalent
to that of Ref.~\onlinecite{DMFT}, after shift in the $k^2$ term
and recognition that the self-consistently determined effective
temperature ($\overline{r}$), depends on the dominant order
$\Phi_{1}$. The reason for equivalence is that in the presence of
only quartic and biquadratic terms, the propagator lines for the
fields $\Phi_{1}$ and $\Phi_{2}$ are continuous and only allow for
self energies with the same field index as for a single field
problem with shifted parameters. In Ref.~\onlinecite{DMFT} the
dynamical mean field theory calculation yielded extensive $S_{c}$
for the single component problem, and hence precisely the same
conclusion is applicable to our model.

The extensive value of $S_{c}$ implies that $N_m\propto e^V$, and
strongly suggests glassiness for $T< T_{K}$ \cite{kirk}.
The condition for possible glassiness formulated
in Refs.~\onlinecite{jorg,DMFT} is that that ratio of the
coherence length to the modulation scale exceeds a number of order
two. As seen from Eq.(\ref{G2}) in our model at low temperatures
$\xi_{2}/l_{2} \ge 2$ satisfying this condition.  Once again,
the realization of the glassy phase depends on the details of
dynamics in a particular measurement, but extensive entropy makes
such an outcome likely.

The high degree of low temperature entropy can be made rigorous.
In all large $n$ (and several Ising) systems \cite{glass}, the extensive
configurational entropy found at higher temperatures by replica
calculations is supplanted by a ground state degeneracy scaling as
the surface area of the system $(S_{ground} \propto q^{d-1}
V^{(d-1)/d})$ \cite{glass},\cite{long} in $d$ spatial dimensions.
By explicit construction, these systems can be shown to possess a
multitude of zero energy domain walls \cite{glass}. These low temperature
excitations going hand in hand with a multiple metastable low energy
states. Numerical
simulations of single component systems in similar clasical
models of liquids also report exceptionally sluggish dynamics
\cite{Grousson,GR} with strong indications of glassiness
\cite{Grousson}. Thus, the non-uniform structures arising in our
model of competing order parameters naturally exhibit slow
dynamics and is likely to become glassy.

\begin{figure}
\includegraphics[width=7cm]{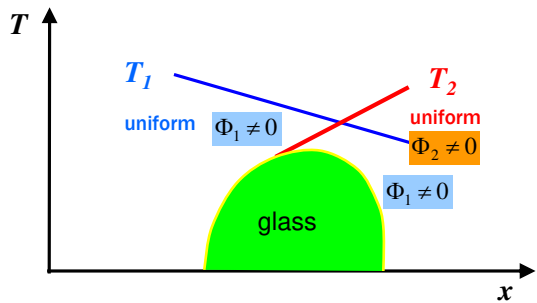}
\caption{Schematic phase diagram beyond the GL theory.
Here we highlight the possibility of a glassy
phase triggered by the competition of two
local orders. Alternatives include the first order
Brazovskii transition into a modulated state, or a transition with
a severely suppressed $T_c$. }
\label{final1}
\end{figure}

Summarizing,the field theoretical analysis accounting for
fluctuations around the inhomogeneous minimizing structure
extends the GL picture and strongly suggests the phase diagram
shown in Fig.(\ref{final1}). For the low $n$ systems of relevance,
the low temperature first order Brazovskii transition can be
pre-empted by a transition into a glass.

\section{Relevance to electronic systems.}

We showed that, when there is a competition between two order
parameters of different origin, and when a general symmetry
allowed gradient-amplitude coupling in a {\it local} theory is
negative ({\it even if of moderate magnitude}), the coexistence of
two orders is inhomogeneous, and, generally, either the dynamics
of the system is slow or a first
order Brazovskii transition occurs. Crucially, even though we
start with a local theory, the inhomogeneous coexistence leads to
a low-energy theory of the same class as considered in models of
self-generated glassiness due to competing length scales of
interaction \cite{jorg,loh}, although the origin of the phenomenon
is very different.

Moreover, the transition temperatures for {\it both} order
parameters are suppressed compared to the mean field value. We
emphasize that the gradient-amplitude coupling is required to
stabilize an inhomogeneous state: in its absence only uniform or
phase-separated configurations are thermodynamically stable, as
has been shown for stripe orders \cite{Pryadko}. Therefore in
competing coexisting phases, $T_c$ is lower than in other parts of
the phase diagram, structure factor measurements will indicate
non-uniform order, and dynamical measurements likely
display slow dynamics. The natural question to ask is what
systems offer the best chance for realization of the model
considered above.

Cuprates provide one obvious example of such competing orders when
static low temperature spin and charge density waves (stripes) are
inhibited in the presence of superconducting order. In these
materials, STM measurements \cite{lang} indicate incommensurate
coexistence of superconductivity and a pseudo-gap state at
nanoscale level; however, the dynamics in this situation is
strongly energy dependent, which suggests that the mapping
on a simple GL theory with temperature-independent coefficients is
insufficient. At least in one example the scaling form of the
dielectric function in the glassy state goes smoothly to quantum
critical scaling as the glass transition temperature tends to zero
\cite{TPark:2005}.

Heavy fermion systems provide perhaps the best chance for
observing the phenomena described here. In materials of the 115
family proximity or coexistence of antiferromagnetic and
superconducting phases is now well established
\cite{TPark:2006,TPark:PNAS}, and experiments indicate an
inhomogeneous coexistence of the two orders in a magnetic
field\cite{MKenzel,Curro3}. Moreover, there is strong evidence
that Cd and Hg dopants \cite{LPham,EBauer} create
antiferromagnetic regions in their vicinity
\cite{Yoshi,Curro1,Curro2}, suggesting that the system is on the
border of inhomogeneous coexistence of two orders. The Neel
temperature drops precipitously if superconducting transition
occurs first \cite{TPark:2006}. No dynamical measurements have yet
been carried out in the relevant regime of the phase diagram, but
it would be interesting to see if, for example, in CeRhIn$_5$
under pressure the spin dynamics as determine by NMR shows
signatures of slowing down or freezing at low temperatures.

In several systems the inhomogeneous coexistence was
proposed in the presence of coupling terms that exist only under
special circumstances \cite{heine,mohamed,Littlewood}. A
particularly relevant example are manganites where the coupling
due to deviation from half-filling that promotes the inhomogeneous
coexistence of the magnetic and charge orders was proposed
recently based on considerations similar to ours
\cite{Littlewood}. As mentioned above, in these materials
glassiness may emerge due to bona fide disorder, and not be
self-generated. Non-trivial couplings appear in some of the
multiferroic materials, e.g. spiral magnets such as RMnO$_{3}$
with R= Tb, Ho, Dy. \cite{multif}

It is important to note that, if we extend the treatment to
include external parameters such as strain and field to act as a
massive (i.e. with fluctuations towards order but no
symmetry breaking since the quadratic coefficient in the GL
expansion remains positive) ``competing orders'' within the GL
framework, the resulting inhomogeneous state only occurs for
moderately large coupling. One
candidate for such a scenario is MnSi, where the low energy theory
exhibiting these features has recently been put forward on the
basis of Dzyaloshinskii-Moriya coupling \cite{Schmalian}.

In conclusion, we believe that many of the observed low
temperature transitions, inhomogeneities, and slow
dynamics/glassiness found in strongly correlated electronic
systems are a natural consequence of competing local orders. As we
illustrated, competing local orders may trigger inhomogeneities
with likely first order transition or possible glassiness. In our
calculations, the proliferation of incommensurate ground and
metastable states is the common origin of both the dramatic
lowering of the transition temperature or viable first order
Brazovskii transition and possible glassy dynamics.

\section{Acknowledgments.} This research was supported by the
US DOE under LDRD X1WX (Z. N. and A. V. B.) and DE-FG02-08ER46492
(I.~V.), and by the CMI of WU (Z. N.).

\end{document}